\documentclass[onecolumn,showpacs,preprintnumbers,amsmath,amssymb]{revtex4}

\usepackage{graphicx}
\usepackage{dcolumn}
\usepackage{bm}

\newcommand{\Real}{\mathop{\textrm{Re}}}

\begin{document}

\preprint{APS/123-QED}

\title{Magnetic-field-induced electric quadrupole moments for relativistic hydrogenlike atoms: Application of the Sturmian expansion of the generalized Dirac--Coulomb Green function}

\author{Patrycja Stefa{\'n}ska}
 \email{pstefanska@mif.pg.gda.pl}
\affiliation{Atomic Physics Division, \mbox{Department of Atomic, Molecular and Optical Physics,} Faculty of Applied Physics and Mathematics, Gda{\'n}sk University of Technology,  Narutowicza 11/12, 80--233 Gda{\'n}sk, Poland
}%


\begin{abstract}
\begin{center}
\textbf{Published as: Phys.\ Rev.\ A 93 (2016) 022504/1-11}
\\*[1ex]
\textbf{doi: 10.1103/PhysRevA.93.022504}\\*[5ex]
\textbf{Abstract} \\*[0.5ex]
\end{center}

We consider a Dirac one-electron atom placed in a weak, static, uniform magnetic field. We show that, to the first order in the strength $B$ of the external field, the only electric multipole moments, which are induced by the perturbation in the atom, are those of an even order. Using the Sturmian expansion of the generalized Dirac--Coulomb Green function [R. Szmytkowski, J. Phys. B \textbf{30}, 825 (1997); \textbf{30}, 2747(E) (1997)], we derive a closed-form expression for the electric quadrupole moment induced in the atom in an arbitrary discrete energy eigenstate. The result, which has the form  of a double finite sum involving the generalized hypergeometric functions ${}_3F_2$ of the unit argument, agrees with the earlier relativistic formula for that quantity, obtained by us for the ground state of the atom.
\end{abstract}

\pacs{31.15.ap, 32.10.Dk, 02.30.Gp}
\maketitle

\section{\label{sec:I}Introduction}
\setcounter{equation}{0}

Calculations of atomic and molecular electromagnetic susceptibilities are undoubtedly one of the most interesting issues of theoretical physics. Analytical investigations of various parameters characterizing a response of a particle to external electric and magnetic fields were carried out already in the early years of quantum mechanics. In Ref.\ \cite{Vlec32}, Van Vleck discussed some properties of \emph{nonrelativistic} hydrogenlike ions, such as the polarizability and the magnetizability. The corresponding \emph{relativistic} analytical calculations were initiated at the beginning of the 1970's by Manakov \emph{et al.}  \cite{Mana74, Zon72}. The main tool used in these calculations has been a Sturmian expansion of the second-order Dirac--Coulomb Green function (DCGF). In 1997, Szmytkowski \cite{Szmy97} proposed such a Sturmian series representation of the first-order DCGF, which turned out to be even more useful in perturbation-theory calculations in relativistic atomic physics. Thus far, it has been successfully used to derive closed-form expressions for many susceptibilities of the Dirac one-electron atom in the ground state, i.e., the static and dynamic electric dipole polarizabilities \cite{Szmy97, Szmy04, Szmy02a}, the induced magnetic anapole moment \cite{Miel06}, the dipole magnetizability \cite{Szmy02b}, the electric and magnetic dipole shielding constants \cite{Szmy11, Stef12}, the magnetic-field-induced electric quadrupole moment \cite{Szmy12}, and the magnetic quadrupole moment induced in the atom by a weak, static, uniform electric field \cite{Szmy14}.

Recently, we have shown that the applicability of this method goes beyond the study of the atomic ground state. In Ref. \cite{Stef15a}, we have analyzed the \emph{magnetic dipole} moment induced in the relativistic hydrogenlike atom by a weak, static, uniform magnetic field. Actually, we derived analytically an expression for the magnetizability of the atom in an arbitrary excited state, which allowed us then to find numerical values of this quantity for a few low-lying discrete energy eigenstates \cite{Stef15b}. In the present work, which is a natural extension of the studies outlined in Ref. \cite{Stef15a}, we present the calculations of the \emph{electric quadrupole} moment induced in an arbitrary discrete energy eigenstate of the Dirac one-electron atom by the same perturbation as in the aforementioned article. In this way, we also generalize the considerations described in Ref.\ \cite{Szmy12}.

The structure of the paper is as follows. In Sec.\ \ref{II}, we present a basic knowledge concerning the relativistic hydrogenlike atom being in an arbitrary discrete energy eigenstate perturbed by a weak, static, uniform magnetic field. Section  \ref{III} provides an analysis of atomic electric multipole moments. We show that in an unperturbed state the only nonvanishing electric multipole moments are those of an even order, and the external magnetic field induces also even-order electric moments only. Therefore, in Sec.\ \ref{IV} we calculate the moment of the lowest possible order, namely the induced electric quadrupole moment, using for this purpose the Sturmian expansion of the first-order DCGF \cite{Szmy97}.

\section{Preliminaries}
\label{II}
\setcounter{equation}{0}

It has been already mentioned in the Introduction that the system to be studied in this paper is the relativistic hydrogenlike atom  with a spinless, pointlike and motionless nucleus of charge $+Ze$, and with an electron of mass $m_e$ and charge $-e$. In the absence of external perturbations, the atomic state energy levels are 
\begin{equation}
E^{(0)} \equiv E_{n \kappa}^{(0)}=m_ec^2\frac{n+\gamma_{\kappa}}{N_{n \kappa}},
\label{2.1}
\end{equation}
with
\begin{equation}
N_{n\kappa}=\sqrt{n^2+2n\gamma_{\kappa}+\kappa^2}
\label{2.2}
\end{equation}
and
\begin{equation}
\gamma_{\kappa}=\sqrt{\kappa^2-(\alpha Z)^2},
\label{2.3}
\end{equation}
where $n$ is the radial quantum number, $\kappa$ is an integer different from zero, while $\alpha$ denotes the Sommerfeld's fine-structure constant. The atomic energy levels from Eq.\ (\ref{2.1}) are degenerate; the normalized to unity eigenfunctions associated with the eigenvalue $E_{n\kappa}^{(0)}$ are
\begin{equation}
\Psi^{(0)}(\boldsymbol{r}) \equiv \Psi_{n \kappa \mu}^{(0)}(\boldsymbol{r}) 
=\frac{1}{r} 
\left(
\begin{array} {c}
P_{n\kappa}^{(0)}(r) \Omega_{\kappa\mu}(\boldsymbol{n}_r) \\ 
\textrm{i} Q_{n\kappa}^{(0)}(r) \Omega_{-\kappa\mu}(\boldsymbol{n}_r) 
\end{array} 
\right).
\label{2.4}
\end{equation}
Here, $\Omega_{\kappa \mu}(\boldsymbol{n}_r)$ (with $\boldsymbol{n}_r=\boldsymbol{r}/r$, $\kappa=\pm1, \pm2, \ldots$ and $\mu=-|\kappa|+\frac{1}{2}, -|\kappa|+\frac{3}{2}, \ldots , |\kappa|-\frac{1}{2}$) are the orthonormal spherical spinors defined as in Ref.\ \cite{Szmy07}, while the radial functions, which are normalized to unity in the sense of
\begin{equation}
\int_0^{\infty} \textrm{d}r 
\left\{
[P_{n\kappa}^{(0)}(r)]^2+[Q_{n\kappa}^{(0)}(r)]^2 
\right\}
=1,
\label{2.5}
\end{equation}
have the form
\begin{equation}
P_{n\kappa}^{(0)}(r)=\tilde{f}_{n\kappa}\sqrt{1+\epsilon_{n\kappa}}
\left(
\lambda_{n\kappa} r\right)^{\gamma_{\kappa}}\textrm{e}^{-\lambda_{n\kappa} r/2}
\left[
L_{n-1}^{(2\gamma_{\kappa})}\left(\lambda_{n\kappa} r\right)
+\frac{\kappa-N_{n\kappa}}{n+2\gamma_{\kappa}}L_{n}^{(2\gamma_{\kappa})}\left(\lambda_{n\kappa} r\right)
\right],
\label{2.6}
\end{equation}
\begin{equation}
Q_{n\kappa}^{(0)}(r)=\tilde{f}_{n\kappa}\sqrt{1-\epsilon_{n\kappa}} 
\left(
\lambda_{n\kappa} r\right)^{\gamma_{\kappa}}\textrm{e}^{-\lambda_{n\kappa} r/2}
\left[
L_{n-1}^{(2\gamma_{\kappa})}\left(\lambda_{n\kappa} r\right)
-\frac{\kappa-N_{n\kappa}}{n+2\gamma_{\kappa}}L_{n}^{(2\gamma_{\kappa})}\left(\lambda_{n\kappa} r\right)
\right],
\label{2.7}
\end{equation}
where $L_{n}^{(\beta)}(\rho)$ is the generalized Laguerre polynomial \cite{Magn66} [it is understood that $L_{-1}^{(2\gamma_{\kappa})}(\rho) \equiv 0$],  
\begin{equation}
\epsilon_{n \kappa}=\frac{E_{n\kappa}^{(0)}}{m_ec^2}=\frac{n+\gamma_{\kappa}}{N_{n\kappa}}, \qquad \lambda_{n\kappa}=\frac{2Z}{a_0 N_{n\kappa}}
\label{2.8}
\end{equation}
with $a_0$ denoting the Bohr radius, and
\begin{equation}
\tilde{f}_{n\kappa}=\sqrt{\frac{Z}{2a_0} \frac{(n+2\gamma_{\kappa}) n!}{N_{n\kappa}^2(N_{n\kappa}-\kappa)\Gamma(n+2\gamma_{\kappa})}}. 
\label{2.9}
\end{equation}

Now, let us assume that the atom is placed in a weak static, uniform magnetic field $\boldsymbol{B}=B\boldsymbol{n}_z$. The energy eigenvalue problem for bound states of such a system is constituted by the Dirac equation
\begin{equation}
\left[
-\mathrm{i}c\hbar\boldsymbol{\alpha}\cdot\boldsymbol{\nabla}
+\frac{1}{2}e c \boldsymbol{\alpha}\cdot \left(\boldsymbol{B} \times \boldsymbol{r}\right)+\beta
m_ec^{2}-\frac{Ze^{2}}{(4\pi\epsilon_{0})r}-E
\right]
\Psi(\boldsymbol{r})=0
\label{2.10}
\end{equation}
(with $\boldsymbol{\alpha}$ and $\beta$ being the standard Dirac matrices \cite{Schi55}) supplemented by the boundary conditions
\begin{equation}
r \Psi(\boldsymbol{r}) \stackrel{r \to 0}{\longrightarrow}0, 
\qquad \qquad 
r^{3/2} \Psi(\boldsymbol{r}) \stackrel{r \to \infty}{\longrightarrow}0. 
\label{2.11}
\end{equation}
Because the functions from Eq.\ (\ref{2.4}) are adjusted to the perturbation $\hat{H}^{(1)}=\frac{1}{2} e c \boldsymbol{\alpha}\cdot\left(\boldsymbol{B}\times \boldsymbol{r}\right)$ appearing in Eq.\ (\ref{2.10}), i.e. they diagonalize the matrix of that perturbation, the solutions of the eigenproblem (\ref{2.10})--(\ref{2.11}), to the first order in $\boldsymbol{B}$, may be approximated as
\begin{equation}
 \Psi(\boldsymbol{r}) \simeq \Psi^{(0)}(\boldsymbol{r})+\Psi^{(1)}(\boldsymbol{r}), \qquad \qquad  E\simeq E^{(0)}+E^{(1)}.
\label{2.12}
\end{equation}
The zeroth-order components $E^{(0)}$ and $\Psi^{(0)}(\boldsymbol{r})$ are given by Eqs.\ (\ref{2.1}) and (\ref{2.4}) (with the space quantization axis chosen along the external magnetic filed direction), respectively, whereas the corrections $\Psi^{(1)}(\boldsymbol{r})$ and $E^{(1)}$ solve the inhomogeneous differential equation
\begin{equation}
\left[
-\textrm{i} c \hbar \boldsymbol{\alpha} \cdot \boldsymbol{\nabla} 
+\beta m_ec^2  
-\frac{Z e^2}{(4\pi \epsilon_0)r} 
-E^{(0)} 
\right]
\Psi^{(1)}(\boldsymbol{r})
=- 
\left[
\frac{1}{2} e c \boldsymbol{{B}} \cdot 
\left(
\boldsymbol{r} \times \boldsymbol{\alpha}
\right)
-E^{(1)} 
\right] 
\Psi^{(0)}(\boldsymbol{r}),
\label{2.13}
\end{equation}
subject to the usual regularity conditions and the orthogonality constraint
\begin{equation}
\int_{\mathbb{R}^3} \textrm{d}^3\boldsymbol{r}\: \Psi^{(0)\dagger}(\boldsymbol{r}) \Psi^{(1)}(\boldsymbol{r})=0.
\label{2.14}
\end{equation}
The integral representation of $\Psi^{(1)}(\boldsymbol{r})$ is
\begin{equation}
\Psi^{(1)}(\boldsymbol{r})=-\frac{1}{2} e c \boldsymbol{B} \cdot \int_{\mathbb{R}^3} \textrm{d}^3\boldsymbol{r'} \:\bar{G}^{(0)}(\boldsymbol{r},\boldsymbol{r}') 
\left( 
\boldsymbol{r}' \times \boldsymbol{\alpha}
\right) 
\Psi^{(0)}(\boldsymbol{r}'),
\label{2.15}
\end{equation}
where $\bar{G}^{(0)}(\boldsymbol{r},\boldsymbol{r}')$ is the generalized Dirac--Coulomb Green function associated with the energy level (\ref{2.1}) of an isolated atom.

\section{Analysis of electric multipole moments of the atom in the magnetic field}
\label{III}
\setcounter{equation}{0}

Let us now consider, which electric multipole moments characterize the electronic cloud of the isolated atom, and which of them may be induced in the system by an external weak, uniform, static magnetic field. The spherical components of the $L$th-order electric multipole moment tensor are defined as
\begin{equation}
\mathcal{Q}_{LM}=\sqrt{\frac{4\pi}{2L+1}} \int_{\mathbb{R}^3} \textrm{d}^3\boldsymbol{r} \: r^L Y_{LM}(\boldsymbol{n}_r) \rho(\boldsymbol{r}),
\label{3.1}
\end{equation}
where $Y_{LM}(\boldsymbol{n}_r)$ is the normalized spherical harmonic defined according to the Condon-Shortley phase convention \cite{Vars75}, while 
\begin{equation}
\rho(\boldsymbol{r})=\frac{-e\Psi^{\dagger}(\boldsymbol{r})\Psi(\boldsymbol{r})}{\int_{\mathbb{R}^3} \textrm{d}^3\boldsymbol{r}'\Psi^{\dagger}(\boldsymbol{r}')\Psi(\boldsymbol{r}')}
\label{3.2}
\end{equation}
is the electronic charge density for the perturbed state $\Psi(\boldsymbol{r})$. Utilizing the first of the approximations from Eq.\ (\ref{2.12}), taking advantage of Eq.\ (\ref{2.14}) and keeping in mind that $\Psi^{(0)}(\boldsymbol{r})$ is normalized to unity, one has
\begin{equation}
\rho(\boldsymbol{r}) \simeq \rho^{(0)}(\boldsymbol{r})+\rho^{(1)}(\boldsymbol{r}),
\label{3.3}
\end{equation}
with
\begin{equation}
\rho^{(0)}(\boldsymbol{r})=-e \Psi^{(0)\dagger}(\boldsymbol{r})\Psi^{(0)}(\boldsymbol{r})
\label{3.4}
\end{equation}
and
\begin{equation}
\rho^{(1)}(\boldsymbol{r})=-e\left[\Psi^{(1)\dagger}(\boldsymbol{r})\Psi^{(0)}(\boldsymbol{r})+\Psi^{(0)\dagger}(\boldsymbol{r})\Psi^{(1)}(\boldsymbol{r})\right].
\label{3.5}
\end{equation}
Consequently, it follows that
\begin{equation}
\mathcal{Q}_{LM} \simeq \mathcal{Q}_{LM}^{(0)}+\mathcal{Q}_{LM}^{(1)},
\label{3.6}
\end{equation}
where
\begin{equation}
\mathcal{Q}_{LM}^{(0)}=-e\sqrt{\frac{4\pi}{2L+1}} \int_{\mathbb{R}^3} \textrm{d}^3\boldsymbol{r} \: \Psi^{(0)\dagger}(\boldsymbol{r}) r^L Y_{LM}(\boldsymbol{n}_r) \Psi^{(0)}(\boldsymbol{r})
\label{3.7}
\end{equation}
is the multipole moment of the isolated atom, while
\begin{equation}
\mathcal{Q}_{LM}^{(1)}=\widetilde{\mathcal{Q}}_{LM}^{(1)}+(-)^{M}\widetilde{\mathcal{Q}}_{L,-M}^{(1)*},
\label{3.8}
\end{equation}
with
\begin{equation}
\widetilde{\mathcal{Q}}_{LM}^{(1)}=-e\sqrt{\frac{4\pi}{2L+1}} \int_{\mathbb{R}^3} \textrm{d}^3\boldsymbol{r} \: \Psi^{(0)\dagger}(\boldsymbol{r}) r^L Y_{LM}(\boldsymbol{n}_r) \Psi^{(1)}(\boldsymbol{r}),
\label{3.9}
\end{equation}
is the first-order correction induced by the perturbing magnetic field. To obtain Eq.\ (\ref{3.8}), we have used the following relation:
\begin{equation}
Y_{LM}(\boldsymbol{n}_r)=(-)^{M}Y_{L,-M}^{*}(\boldsymbol{n}_r).
\label{3.10}
\end{equation}

To analyze the electric moments of the unperturbed atom, in the first step we put Eq.\ (\ref{2.4}) into Eq.\ (\ref{3.7}), yielding
\begin{eqnarray}
\mathcal{Q}_{LM}^{(0)}=-e\sqrt{\frac{4\pi}{2L+1}}
\left\{
\left<\Omega_{\kappa \mu}|Y_{LM} \Omega_{\kappa \mu}\right> \int_0^{\infty} \textrm{d}r \: r^L \left[P_{n \kappa}^{(0)}(r)\right]^2 
+\left<\Omega_{-\kappa \mu}|Y_{LM} \Omega_{-\kappa \mu}\right> \int_0^{\infty} \textrm{d}r \: r^L \left[Q_{n \kappa}^{(0)}(r)\right]^2
\right\},
\label{3.11}
\end{eqnarray}
where the shorthand bracket notation
\begin{equation}
\left<\Omega_{\kappa \mu}|Y_{LM} \Omega_{\kappa' \mu'}\right> \equiv \oint_{4\pi}\mathrm{d}^{2}\boldsymbol{n}_{r}\:
\Omega_{\kappa\mu}^{\dag}(\boldsymbol{n}_{r})
Y_{LM}(\boldsymbol{n}_{r})\Omega_{\kappa'\mu'}(\boldsymbol{n}_{r})
\label{3.12}
\end{equation}
has been used for the angular integrals. Exploiting the identity  \cite[Eq.\ (3.1.3)]{Szmy07}
\begin{equation}
\boldsymbol{n}_{r}\cdot\boldsymbol{\sigma}
\Omega_{\kappa\mu}(\boldsymbol{n}_{r})
=-\Omega_{-\kappa\mu}(\boldsymbol{n}_{r}),
\label{3.13}
\end{equation}
it is easy to prove that
\begin{equation}
\left<\Omega_{-\kappa \mu}|Y_{LM} \Omega_{-\kappa' \mu'}\right>=\left<\Omega_{\kappa \mu}|Y_{LM} \Omega_{\kappa' \mu'}\right>,
\label{3.14}
\end{equation}
and so Eq.\ (\ref{3.11}) may be cast into the form
\begin{equation}
\mathcal{Q}_{LM}^{(0)}=-e\sqrt{\frac{4\pi}{2L+1}} \left<\Omega_{\kappa \mu}|Y_{LM} \Omega_{\kappa \mu}\right> \int_0^{\infty} \textrm{d}r \: r^L \left\{\left[P_{n \kappa}^{(0)}(r)\right]^2 +\left[Q_{n \kappa}^{(0)}(r)\right]^2 \right\}.
\label{3.15}
\end{equation}
The angular integral from the above equation may be conveniently evaluated from the general formula
\begin{eqnarray}
\sqrt{\frac{4\pi}{2L+1}}\left<\Omega_{\kappa \mu}|Y_{LM} \Omega_{\kappa' \mu'}\right> = (-)^{\mu+1/2} 2\sqrt{|\kappa \kappa'|}
\left(
\begin{array}{ccc}
|\kappa|-\frac{1}{2} & L & |\kappa'|-\frac{1}{2} \\
-\frac{1}{2} & 0 & \frac{1}{2} \\
\end{array}
\right)
\left(
\begin{array}{ccc}
|\kappa|-\frac{1}{2} & L & |\kappa'|-\frac{1}{2} \\
\mu & M & -\mu' \\
\end{array}
\right)
\Pi(l_{\kappa}, L, l_{\kappa'}), \quad
\label{3.16}
\end{eqnarray}
where
\begin{equation}
\Pi(l_{\kappa}, L, l_{\kappa'})=
\left\{
\begin{array}{lcl}
1 & \textrm{for} & l_{\kappa}+L+l_{\kappa'} \: \textrm{even} \\
0 & \textrm{for} & l_{\kappa}+L+l_{\kappa'} \: \textrm{odd},
\end{array}
\right.
\label{3.17}
\end{equation}
with $l_{\kappa}=|\kappa+\frac{1}{2}|-\frac{1}{2}$ (and similarly for $l_{\kappa'}$), whereas {$\displaystyle\left(\begin{array}{ccc} j_{a} & j_{b} & j_{c} \\
m_{a} & m_{b} & m_{c}\end{array} \right)$} denotes the Wigner's 3$j$ coefficient. Exploiting the selection rules expressed in Eq.\ (\ref{3.17}) and some basic properties of the 3$j$ coefficients, one infers that the only cases when $\mathcal{Q}_{LM}^{(0)}$ does not vanish are those with $M=0$ and with $L$ being an even number satisfying the inequality $0 \leqslant L \leqslant 2|\kappa|$. From the physical point of view, this means that the atom being in the state characterized by the quantum number $\kappa$ has  the permanent electric multipole moments only of an even order, up to $2|\kappa|$ inclusive.

Next, we shall focus our interest on the first-order induced multipole moments $\mathcal{Q}_{LM}^{(1)}$. According to the relation in Eq.\ (\ref{3.8}), for convenience, now we will consider the component $\widetilde{\mathcal{Q}}_{LM}^{(1)}$. Therefore, we insert Eq.\ (\ref{2.15}) into Eq.\ (\ref{3.9}) and obtain 
\begin{equation}
\widetilde{\mathcal{Q}}_{LM}^{(1)}=\frac{1}{2} \sqrt{\frac{4\pi}{2L+1}} e^2 c B \int_{\mathbb{R}^3} \textrm{d}^3\boldsymbol{r} \int_{\mathbb{R}^3} \textrm{d}^3\boldsymbol{r}' \: \Psi^{(0)\dagger}(\boldsymbol{r}) r^L Y_{LM}(\boldsymbol{n}_r) \bar{G}^{(0)}(\boldsymbol{r},\boldsymbol{r}') \boldsymbol{n}_z \cdot (\boldsymbol{r}' \times \boldsymbol{\alpha}) \Psi^{(0)}(\boldsymbol{r}').
\label{3.18}
\end{equation}
To be able to analyze the induced moments as in the case of the permanent moments, we have to invoke the partial-wave expansion of the generalized Dirac--Coulomb Green function, which is
{\small
\begin{eqnarray}
\bar{G}\mbox{}^{(0)}(\boldsymbol{r},\boldsymbol{r}')
= \frac{4\pi\epsilon_{0}}{e^{2}} 
\sum_{\substack{\kappa'=-\infty \\ (\kappa'\neq0)}}^{\infty}
\sum_{\mu'=-|\kappa'|+1/2}^{|\kappa'|-1/2}\frac{1}{rr'} 
\left(
\begin{array}{cc}
\bar{g}\mbox{}^{(0)}_{(++)\kappa'}(r,r')
\Omega_{\kappa' \mu'}(\boldsymbol{n}_{r})
\Omega_{\kappa' \mu'}^{\dag}(\boldsymbol{n}_{r}^{\prime}) &
-\mathrm{i}\bar{g}\mbox{}^{(0)}_{(+-)\kappa'}(r,r')
\Omega_{\kappa' \mu'}(\boldsymbol{n}_{r})
\Omega_{-\kappa' \mu'}^{\dag}(\boldsymbol{n}_{r}^{\prime}) \\
\mathrm{i}\bar{g}\mbox{}^{(0)}_{(-+)\kappa'}(r,r')
\Omega_{-\kappa' \mu'}(\boldsymbol{n}_{r})
\Omega_{\kappa' \mu'}^{\dag}(\boldsymbol{n}_{r}^{\prime}) &
\bar{g}\mbox{}^{(0)}_{(--)\kappa'}(r,r')
\Omega_{-\kappa' \mu'}(\boldsymbol{n}_{r})
\Omega_{-\kappa' \mu'}^{\dag}(\boldsymbol{n}_{r}^{\prime}) 
\end{array} 
\right).
\nonumber \\
\label{3.19}%
\end{eqnarray}}
Plugging Eqs.\ (\ref{2.4}) and (\ref{3.19}) into Eq.\ (\ref{3.18}), then exploiting the identity from Eq.\ (\ref{3.14}) and the following relation:
\begin{equation}
\left<\Omega_{-\kappa \mu}|\boldsymbol{n}_{z} \cdot (\boldsymbol{n}_{r} \times \boldsymbol{\sigma})\Omega_{\kappa' \mu'}\right>=-\left<\Omega_{\kappa \mu}|\boldsymbol{n}_{z} \cdot (\boldsymbol{n}_{r} \times \boldsymbol{\sigma})\Omega_{-\kappa' \mu'}\right>
\label{3.20}
\end{equation}
(which can be easily proved with the use of some properties of the Pauli matrices), we arrive at
\begin{equation}
\widetilde{\mathcal{Q}}_{LM}^{(1)}=\frac{\textrm{i}}{2} \sqrt{\frac{4\pi}{2L+1}} (4\pi \epsilon_0) c B 
\sum_{\kappa' \mu'} R_{\kappa'}^{L 1} \left<\Omega_{\kappa' \mu'}|\boldsymbol{n}_{z} \cdot (\boldsymbol{n}_{r} \times \boldsymbol{\sigma})\Omega_{-\kappa \mu}\right> \left<\Omega_{\kappa \mu}|Y_{LM}\Omega_{\kappa' \mu'}\right>,
\label{3.21}
\end{equation}
where we define
\begin{equation}
R_{\kappa'}^{LL'}=\int_0^{\infty} \textrm{d}r \int_0^{\infty} \textrm{d}r'   
\left(
\begin{array}{cc} 
P_{n\kappa}^{(0)}(r) &  
Q_{n\kappa}^{(0)}(r)    
\end{array}\right)  
r^L \bar{\mathsf{G}}_{\kappa'}^{(0)}(r,r')  r'^{L'}  
\left(
\begin{array}{c}
 Q_{n\kappa}^{(0)}(r') \\  
P_{n\kappa}^{(0)}(r')
\end{array}
\right),
\label{3.22}
\end{equation}
with
\begin{equation}
\bar{\mathsf{G}}\mbox{}^{(0)}_{\kappa'}(r,r')
=\left(
\begin{array}{cc}
\bar{g}\mbox{}^{(0)}_{(++)\kappa'}(r,r') &
\bar{g}\mbox{}^{(0)}_{(+-)\kappa'}(r,r') \\*[1ex]
\bar{g}\mbox{}^{(0)}_{(-+)\kappa'}(r,r') &
\bar{g}\mbox{}^{(0)}_{(--)\kappa'}(r,r')
\end{array}
\right)
\label{3.23}
\end{equation}
being the radial generalized Dirac--Coulomb Green function associated with the combined total angular momentum and parity quantum number $\kappa'$.

To tackle the first angular integral appearing on the right-hand side of Eq.\ (\ref{3.21}), we shall exploit the relation, Eq.\ (3.1.6) in Ref.\ \cite{Szmy07},
\begin{eqnarray}
\boldsymbol{n}_z\cdot 
\left( 
\boldsymbol{n}_r \times \boldsymbol{\sigma} 
\right) 
\Omega_{\kappa \mu}(\boldsymbol{n}_r)=
\textrm{i}\frac{4\mu \kappa}{4\kappa^2-1}\Omega_{-\kappa \mu}(\boldsymbol{n}_r)
+\textrm{i}\frac{\sqrt{\left(\kappa+\frac{1}{2}\right)^2-\mu^2}}{|2\kappa+1|}\Omega_{\kappa+1, \mu}(\boldsymbol{n}_r)
-\textrm{i}\frac{\sqrt{\left(\kappa-\frac{1}{2}\right)^2-\mu^2}}{|2\kappa-1|}\Omega_{\kappa-1, \mu}(\boldsymbol{n}_r) \quad
\label{3.24} 
\end{eqnarray}
and the fact, that the spherical spinors form the orthonormal set on the unit sphere, i.e.:
\begin{equation}
\oint_{4\pi} \textrm{d}^2\boldsymbol{n}_r \: \Omega_{\kappa \mu}^{\dagger}(\boldsymbol{n}_r) \Omega_{\kappa' \mu'}(\boldsymbol{n}_r)=\delta_{\kappa \kappa'}\delta_{\mu \mu'}.
\label{3.25}
\end{equation}
After these steps, we obtain
\begin{eqnarray}
\widetilde{\mathcal{Q}}_{LM}^{(1)}&=& \sqrt{\frac{4\pi}{2L+1}} (4\pi \epsilon_0) c B \sum_{\kappa'} R_{\kappa'}^{L 1} \left<\Omega_{\kappa\mu}|Y_{LM}\Omega_{\kappa'\mu} \right> 
{}
\nonumber \\
&& \times 
\left[ 
\frac{2\kappa \mu}{4\kappa^2-1} \delta_{\kappa',\kappa} - \frac{\sqrt{\left(\kappa-\frac{1}{2} \right)^2-\mu^2}}{2|2\kappa-1|} \delta_{\kappa',-\kappa+1}+ \frac{\sqrt{\left(\kappa+\frac{1}{2} \right)^2-\mu^2}}{2|2\kappa+1|} \delta_{\kappa',-\kappa-1} 
\right].
\label{3.26}
\end{eqnarray}
Due to the formula in Eq.\ (\ref{3.16}), the remaining angular integrals (corresponding to each of the symmetries determined by the Kronecker's deltas) can be  expressed in terms of the Wigner's 3$j$ coefficients. Utilizing the selection rules embodied in Eq.\ (\ref{3.17}) and properties of the 3$j$ coefficients, one deduces that in all the three cases $\widetilde{\mathcal{Q}}_{LM}^{(1)}$ does not vanish if and only if $M=0$ and $L$ is an even number, such that $2 \leqslant L \leqslant 2|\kappa|$. In other words, in a given $\kappa$-state of the Dirac one-electron atom a weak, uniform, static magnetic field induces only even-order electric multipole moments, with $2 \leqslant L \leqslant 2|\kappa|$. A detailed analysis of the induced electric \emph{quadrupole} moment ($L=2$), which is the most interesting one, will be carried out in the next section.

\section{Evaluation of the induced electric quadrupole moment}
\label{IV}
\setcounter{equation}{0}

Before calculating the induced electric quadrupole moment, it will be natural to determine the permanent moment of the atom. For this purpose, we shall insert $L=2$ and $M=0$ into Eq.\ (\ref{3.15}) and consider the angular and the radial parts separately. To tackle the angular integral, we will exploit Eq.\ (\ref{3.16}) and utilize some basic properties and the following expression for the 3$j$ coefficients \cite{Edmo57}:
\begin{equation}
\left(
\begin{array}{ccc}
j & j & 2 \\
m & -m & 0
\end{array}
\right)
=(-)^{j-m} \frac{2[3m^2-j(j+1)]}{\sqrt{(2j+3)(2j+2)(2j+1)2j(2j-1)}}.
\label{4.1}
\end{equation}
In turn, to determine the radial part of the formula (\ref{3.15}), we shall use Eqs.\ (\ref{2.6})--(\ref{2.9}) and take into account the recurrence formula for the Laguerre polynomials, Eq.\ (8.971.5) in Ref.\ \cite{Grad94},
\begin{equation}
L_n^{(\beta)}(\rho)=L_n^{(\beta+1)}(\rho)-L_{n-1}^{(\beta+1)}(\rho)
\label{4.2}
\end{equation}
and the orthogonality relation, Eq.\ (7.414.3) in Ref.\ \cite{Grad94}, 
\begin{equation}
\int_0^{\infty} \textrm{d}\rho \: \rho^{\beta}\textrm{e}^{-\rho}L_m^{(\beta)}(\rho)L_n^{(\beta)}(\rho) 
=\frac{\Gamma(n+\beta+1)}{n!}\delta_{m n} 
\qquad \qquad 
[\Real \beta>-1].
\label{4.3}
\end{equation} 
Combining the partial results, as shown in Eq.\ (\ref{3.15}), we obtain
\begin{eqnarray}
\mathcal{Q}_{2 0}^{(0)}=-\frac{e a_0^2}{Z^2}  \frac{4\kappa^2-12\mu^2-1}{8(4\kappa^2-1)}  
\left[
5(n+\gamma_{\kappa})^4+(3\kappa^2-6\gamma_{\kappa}^2+1)(n+\gamma_{\kappa})^2-3\kappa (n+\gamma_{\kappa}) N_{n \kappa}+(\gamma_{\kappa}^2-\kappa^2)(\gamma_{\kappa}^2-1)
\right].
\label{4.4}
\end{eqnarray}

We turn now to the derivation of the expression for the induced electric quadrupole moment $\mathcal{Q}_{20}^{(1)} \equiv \mathcal{Q}^{(1)}$. At first we attack the angular integrals appearing in Eq.\ (\ref{3.26}). According to the formula given in Eq.\ (\ref{3.16}), they can be evaluated using the 3$j$ coefficients. In this special case, i.e. when $L=2$ and $M=0$, an extremely useful relation is the following one \cite{Edmo57}:
\begin{equation}
\left(
\begin{array}{ccc}
j+1 & j & 2 \\
m & -m & 0
\end{array}
\right)
=(-)^{j-m+1} 2m \sqrt{\frac{6(j+m+1)(j-m+1)}{(2j+4)(2j+3)(2j+2)(2j+1)2j}}.
\label{4.5}
\end{equation}
Utilizing Eqs.\ (\ref{3.16}) and (\ref{4.5}) in Eq.\ (\ref{3.26}), having regarded the identity (\ref{3.8}), we arrive at
\begin{eqnarray}
\mathcal{Q}^{(1)}= \frac{(4\pi \epsilon_0) c B \mu}{4\kappa^2-1} \sum_{\kappa'} 
\left[ 
\frac{\kappa(4\kappa^2-12\mu^2-1)}{4\kappa^2-1} \delta_{\kappa',\kappa}
+\frac{3\left[(2\kappa-1)^2-4\mu^2\right]}{2(2\kappa-1)(2\kappa-3)} \delta_{\kappa',-\kappa+1}
\frac{3\left[(2\kappa+1)^2-4\mu^2\right]}{2(2\kappa+1)(2\kappa+3)} \delta_{\kappa',-\kappa-1}
\right] R_{\kappa'}^{L 1}.
\nonumber \\
\label{4.6}
\end{eqnarray}
The expression for $R_{\kappa'}^{L 1}$ given by Eq.\ (\ref{3.22}) contains the generalized radial Dirac--Coulomb Green function. Because the Sturmian expansion of that function depends on the relationship between numbers $\kappa$ and $\kappa'$ \cite{Szmy97}:
\begin{equation}
\bar{\mathsf{G}}_{\kappa'}^{(0)}(r,r')
=
\sum_{n'=-\infty}^{\infty}{\frac{1}{\mu_{n'\kappa'}^{(0)}-1}}
\left( 
\begin{array}{c}
S_{n'\kappa'}^{(0)}(r) \\
T_{n'\kappa'}^{(0)}(r)
\end{array} 
\right) 
\left(
\begin{array}{cc} 
\mu_{n'\kappa'}^{(0)}S_{n'\kappa'}^{(0)}(r') & 
T_{n'\kappa'}^{(0)}(r')
\end{array}
\right) \qquad \qquad (\textrm{for} \quad \kappa' \neq \kappa),
\label{4.7}
\end{equation}
\begin{eqnarray}
\bar{\mathsf{G}}_{\kappa}^{(0)}(r,r')
&=&
\sum_{\substack{n'=-\infty\\(n'\neq n)}}^{\infty}{\frac{1}{\mu_{n' \kappa}^{(0)}-1}}
\left( 
\begin{array}{c}
S_{n'\kappa}^{(0)}(r) \\
T_{n'\kappa}^{(0)}(r)
\end{array} 
\right)
\left( 
\begin{array}{cc} 
\mu_{n'\kappa}^{(0)} S_{n'\kappa}^{(0)}(r') & 
T_{n'\kappa}^{(0)}(r')\end{array}\right)
 +
\left(
\epsilon_{n\kappa}-\frac{1}{2} 
\right)
\left( 
\begin{array}{c}
S_{n\kappa}^{(0)}(r) \\
T_{n\kappa}^{(0)}(r)
\end{array} 
\right)
\left( 
\begin{array}{cc} 
S_{n\kappa}^{(0)}(r') & 
T_{n\kappa}^{(0)}(r') 
\end{array} 
\right) 
\nonumber \\
&& +
\left( 
\begin{array}{c}
I_{n\kappa}^{(0)}(r) \\
K_{n\kappa}^{(0)}(r)
\end{array} 
\right)
\left( 
\begin{array}{cc} 
S_{n\kappa}^{(0)}(r') &  
T_{n\kappa}^{(0)}(r') 
\end{array} \right)
 +
\left( 
\begin{array}{c}
S_{n\kappa}^{(0)}(r) \\
T_{n\kappa}^{(0)}(r)
\end{array} 
\right)
\left( 
\begin{array}{cc} 
J_{n\kappa}^{(0)}(r') & 
K_{n\kappa}^{(0)}(r')
\end{array} 
\right) \qquad  (\textrm{for} \quad \kappa'=\kappa),
\label{4.8}
\end{eqnarray}
it will be justified to rewrite Eq.\ (4.6) in the following form:
\begin{equation}
\mathcal{Q}^{(1)}=\mathcal{Q}_{\kappa}^{(1)}+\mathcal{Q}_{-\kappa+1}^{(1)}+\mathcal{Q}_{-\kappa-1}^{(1)}
\label{4.9}
\end{equation} 
and consider the first term on the right-hand side of above equation separately from the sum of the other two components. In Eqs.\ (\ref{4.7})--(\ref{4.8})
\begin{equation}
S_{n'\kappa'}^{(0)}(r)=\tilde{g}_{n'\kappa'}\sqrt{1+\epsilon_{n\kappa}}
\left(\lambda_{n\kappa} r\right)^{\gamma_{\kappa'}} 
\textrm{e}^{-\lambda_{n\kappa} r/2}  
\left[
L_{|n'|-1}^{(2\gamma_{\kappa'})}\left(\lambda_{n\kappa} r\right) 
+\frac{\kappa'-N_{n'\kappa'}}{|n'|+2\gamma_{\kappa'}}L_{|n'|}^{(2\gamma_{\kappa'})}\left(\lambda_{n\kappa} r\right)
\right]
\label{4.10}
\end{equation}
and
\begin{equation}
T_{n'\kappa'}^{(0)}(r)=\tilde{g}_{n'\kappa'}\sqrt{1-\epsilon_{n\kappa}}
\left(\lambda_{n\kappa} r\right)^{\gamma_{\kappa'}} 
\textrm{e}^{-\lambda_{n\kappa} r/2}  
\left[
L_{|n'|-1}^{(2\gamma_{\kappa'})}\left(\lambda_{n\kappa} r\right) 
-\frac{\kappa'-N_{n'\kappa'}}{|n'|+2\gamma_{\kappa'}}L_{|n'|}^{(2\gamma_{\kappa'})}\left(\lambda_{n\kappa} r\right)
\right],
\label{4.11}
\end{equation}
with
\begin{equation}
\tilde{g}_{n'\kappa'}=\sqrt{\frac{N_{n \kappa}(|n'|+2\gamma_{\kappa'})|n'|!}{2 Z N_{n'\kappa'}(N_{n'\kappa'}-\kappa')\Gamma(|n'|+2\gamma_{\kappa'})}}   
\label{4.12}
\end{equation}
are the radial Dirac--Coulomb Sturmian functions associated with the hydrogenic discrete state energy level $E_{n\kappa}^{(0)}$, and
\begin{equation}
\mu_{n'\kappa'}^{(0)}=\frac{|n'|+\gamma_{\kappa'}+N_{n'\kappa'}}{n+\gamma_{\kappa}+N_{n \kappa}},
\label{4.13}
\end{equation}
where
\begin{equation}
N_{n'\kappa'}
=\pm\sqrt{\left(|n'|+\gamma_{\kappa'}\right)^2+(\alpha Z)^2}
=\pm \sqrt{|n'|+2|n'|\gamma_{\kappa'}+\kappa'^2}
\label{4.14}
\end{equation}
is a so-called apparent principal quantum number, which assumes the positive values for $n'>0$ and negative for $n'<0$; for $n'=0$, in the definition (\ref{4.14}) one chooses the plus sign if $\kappa'<0$ and the minus sign if $\kappa'>0$. Moreover, the functions $I_{n\kappa}^{(0)}(r)$, $J_{n\kappa}^{(0)}(r)$ and $K_{n\kappa}^{(0)}(r)$, also appearing in Eq.\ (\ref{4.8}), are defined as \cite{Szmy97}
\begin{equation}
I_{n\kappa}^{(0)}(r)
=\epsilon_{n \kappa} 
\left[
-\omega_{n\kappa}^{(+)}S_{n\kappa}^{(0)}(r)+\xi_{n\kappa}^{(+)}(r)T_{n\kappa}^{(0)}(r)
\right], 
\label{4.15}
\end{equation}
\begin{equation}
J_{n\kappa}^{(0)}(r)
=\epsilon_{n \kappa} 
\left[
-\omega_{n\kappa}^{(-)}S_{n\kappa}^{(0)}(r)+\xi_{n\kappa}^{(+)}(r)T_{n\kappa}^{(0)}(r)
\right], 
\label{4.16}
\end{equation}
\begin{equation}
K_{n\kappa}^{(0)}(r)
=\epsilon_{n \kappa} 
\left[
\xi_{n\kappa}^{(-)}(r)S_{n\kappa}^{(0)}(r)+\omega_{n\kappa}^{(+)}T_{n\kappa}^{(0)}(r)
\right], 
\label{4.17}
\end{equation}
where $\omega_{n\kappa}^{(\pm)}=\kappa \pm (2\epsilon_{n\kappa})^{-1}$ and $\xi_{n\kappa}^{(\pm)}(r)=m_{e} c (1 \pm \epsilon_{n\kappa}) r/\hbar \pm \alpha Z $.

Now, let us focus on the first component of the induced quadrupole moment, i.e. $\mathcal{Q}_{\kappa}^{(1)}$. In view of Eqs.\ (\ref{3.22}), (\ref{4.6}), (\ref{4.9}) and the formula (\ref{4.8}) for the Green function, suitable in this case, it will be convenient to write it in the following form:
\begin{equation}
\mathcal{Q}_{\kappa}^{(1)}=(4\pi\epsilon_0) c B \kappa \mu \frac{4\kappa^2-12\mu^2-1}{(4\kappa^2-1)^2} 
\left[
\mathcal{I}_{\kappa}^{(\infty)}+\mathcal{I}_{\kappa}^{(a)}+\mathcal{I}_{\kappa}^{(b)}+\mathcal{I}_{\kappa}^{(c)} 
\right],
\label{4.18}
\end{equation}
with the components
\begin{eqnarray}
\mathcal{I}_{\kappa}^{(\infty)}=\sum_{\substack{{n'=-\infty}\\(n' \neq n)}}^{\infty}{\frac{1}{\mu_{n' \kappa}^{(0)}-1}} 
\int_0^{\infty} \textrm{d}r \: r^2  
\left[
P_{n\kappa}^{(0)}(r) S_{n'\kappa}^{(0)}(r)+ Q_{n\kappa}^{(0)}(r) T_{n'\kappa}^{(0)}(r)
\right] 
\nonumber \\
\times
\int_0^{\infty} \textrm{d}r' \: r' 
\left[
\mu_{n' \kappa}^{(0)} Q_{n\kappa}^{(0)}(r') S_{n'\kappa}^{(0)}(r')+ P_{n\kappa}^{(0)}(r') T_{n'\kappa}^{(0)}(r')
\right],
\label{4.19}
\end{eqnarray}
\begin{eqnarray}
\mathcal{I}_{\kappa}^{(a)}= 
\left(\epsilon_{n\kappa}-\frac{1}{2} \right) 
\int_0^{\infty} \textrm{d}r \: r^2 
\left[
P_{n\kappa}^{(0)}(r) S_{n\kappa}^{(0)}(r)+ Q_{n\kappa}^{(0)}(r) T_{n\kappa}^{(0)}(r)
\right] 
\int_0^{\infty} \textrm{d}r' \: r' 
\left[
Q_{n\kappa}^{(0)}(r') S_{n\kappa}^{(0)}(r')+ P_{n\kappa}^{(0)}(r') T_{n\kappa}^{(0)}(r')
\right], \quad
\label{4.20}
\end{eqnarray}
\begin{eqnarray}
\mathcal{I}_{\kappa}^{(b)}=  
\int_0^{\infty} \textrm{d}r \: r^2 
\left[
P_{n\kappa}^{(0)}(r) I_{n\kappa}^{(0)}(r)+ Q_{n\kappa}^{(0)}(r) K_{n\kappa}^{(0)}(r)
\right]
\int_0^{\infty} \textrm{d}r' \: r' 
\left[
Q_{n\kappa}^{(0)}(r') S_{n\kappa}^{(0)}(r')+ P_{n\kappa}^{(0)}(r') T_{n\kappa}^{(0)}(r')
\right], 
\label{4.21}
\end{eqnarray}
\begin{eqnarray}
\mathcal{I}_{\kappa}^{(c)}=
\int_0^{\infty} \textrm{d}r \: r^2 
\left[
P_{n\kappa}^{(0)}(r) S_{n\kappa}^{(0)}(r)+ Q_{n\kappa}^{(0)}(r) T_{n\kappa}^{(0)}(r)
\right] 
\int_0^{\infty} \textrm{d}r' \: r' 
\left[
Q_{n\kappa}^{(0)}(r') J_{n\kappa}^{(0)}(r')+ P_{n\kappa}^{(0)}(r') K_{n\kappa}^{(0)}(r')
\right]. 
\label{4.22}
\end{eqnarray}
Making use of Eqs.\ (\ref{4.15})--(\ref{4.17}) and the relations
\begin{equation}
S_{n\kappa}^{(0)}(r)=\frac{\sqrt{a_0}N_{n\kappa}}{Z}P_{n\kappa}^{(0)}(r), \qquad T_{n\kappa}^{(0)}(r)=\frac{\sqrt{a_0}N_{n\kappa}}{Z}Q_{n\kappa}^{(0)}(r),
\label{4.23}
\end{equation}
after some algebra, one can prove that
\begin{equation}
\mathcal{I}_{\kappa}^{(a)}+\mathcal{I}_{\kappa}^{(c)}=0
\label{4.24}
\end{equation}
and 
\begin{eqnarray}
\mathcal{I}_{\kappa}^{(b)}=-\frac{a_0 N_{n\kappa}^2}{Z^2} \int_0^{\infty} \textrm{d}r' \: r' P_{n\kappa}^{(0)}(r')Q_{n\kappa}^{(0)}(r') 
\left\{
\int_0^{\infty}  \textrm{d}r \: r^2 \left([P_{n\kappa}^{(0)}(r)]^2-[Q_{n\kappa}^{(0)}(r)]^2 \right) \right.
\nonumber \\
\left.
+3\epsilon_{n \kappa} \int_0^{\infty} \textrm{d}r \: r^2 \left( [P_{n\kappa}^{(0)}(r)]^2+[Q_{n\kappa}^{(0)}(r)]^2 \right)
\right\}.
\label{4.25}
\end{eqnarray}
Utilizing Eqs.\ (\ref{2.6})--(\ref{2.9}) and the relations (\ref{4.2})--(\ref{4.3}) satisfying by the Laguerre polynomials, after some rearrangements the above formula becomes
\begin{eqnarray}
\mathcal{I}_{\kappa}^{(b)}=\frac{\alpha a_0^4}{4 Z^4} (n+\gamma_{\kappa}) \left[N_{n\kappa}-2\kappa(n+\gamma_{\kappa})\right]\left[\left(10n^2+20n\gamma_{\kappa}-3\kappa^2+7\gamma_{\kappa}^2+5\right)N_{n\kappa}^2 
-3(n+\gamma_{\kappa})(n+\gamma_{\kappa}+2\kappa N_{n\kappa})\right]. \quad
\label{4.26}
\end{eqnarray}
To find the expression for $R_{\kappa}^{(\infty)}$, we put Eqs.\ (\ref{2.6})--(\ref{2.9}) and (\ref{4.10})--(\ref{4.13}) into Eq.\ (\ref{4.19}) and exploit the relations (\ref{4.2})--(\ref{4.3}). This gives
\begin{eqnarray}
\mathcal{I}_{\kappa}^{(\infty)}=
\frac{\alpha a_0^4}{Z^4} 
\frac{N_{n \kappa}^3\Gamma(n+2\gamma_{\kappa}+1)}{64n!(N_{n \kappa}-\kappa)}
\sum_{\substack{{n'=-\infty}\\(n' \neq n)}}^{\infty}\frac{|n'|!\left(A_{n'}^{(1)}\delta_{|n'|,n-2}\!+\!A_{n'}^{(2)}\delta_{|n'|,n-1}\!+\!A_{n'}^{(3)}\delta_{|n'|,n}\!+\!A_{n'}^{(4)}\delta_{|n'|,n+1}\!+\!A_{n'}^{(5)}\delta_{|n'|,n+2} 
\right)}
{N_{n' \kappa} (N_{n' \kappa}-\kappa)\Gamma(|n'|+2\gamma_{\kappa}+1)},
\nonumber \\
\label{4.27}
\end{eqnarray}
with the coefficients
\begin{equation}
A_{n'}^{(1)}=n^2(n-1)^2 \left(N_{n'\kappa}-\kappa\right) \left\{(n-2)(n+2\gamma_{\kappa}-2)+\left(N_{n'\kappa}-\kappa \right) \left[4(n+\gamma_{\kappa}-1)\epsilon_{n\kappa}-\kappa+N_{n\kappa}\right]\right\},
\label{4.28}
\end{equation}
\begin{eqnarray}
A_{n'}^{(2)}&=&n^2 
\left\{
2(N_{n'\kappa}-\kappa)\left[2(n+\gamma_{\kappa})(N_{n\kappa}-\kappa)+\epsilon_{n\kappa}\Delta_{n\kappa}^{(-)}\right] 
+(n-1)(n+2\gamma_{\kappa}-1)\left[4(n+\gamma_{\kappa}-1)+\epsilon_{n\kappa}(N_{n\kappa}-\kappa)\right]
\right\}
\nonumber \\
&& \times
\left\{
(N_{n\kappa}+N_{n'\kappa}) \left[(N_{n\kappa}-\kappa)(N_{n'\kappa}-\kappa)-(n-1)(n+2\gamma_{\kappa}-1)\right] +(N_{n'\kappa}-\kappa)(2n+2\gamma_{\kappa}-1)
\right\},
\label{4.29}
\end{eqnarray}
\begin{eqnarray}
A_{n'}^{(3)}&=&2\left(N_{n\kappa}-\kappa\right)^2 
\left\{
\epsilon_{n\kappa} \left[2(n+\gamma_{\kappa})(2\kappa+N_{n\kappa}-N_{n'\kappa}) \right]+(N_{n\kappa}+\kappa)(N_{n\kappa}-N_{n'\kappa})
\right\}
\nonumber \\
&&\times
\left\{
2(n+\gamma_{\kappa})(N_{n\kappa}+\kappa)[\epsilon_{n\kappa}(2\kappa-N_{n\kappa}-N_{n'\kappa})+3]-(N_{n\kappa}+N_{n'\kappa})\Delta_{n\kappa}^{(+)}
\right\},
\label{4.30}
\end{eqnarray}
\begin{eqnarray}
A_{n'}^{(4)}=-(n+2\gamma_{\kappa}+1)^2(N_{n\kappa}-\kappa)^2 \left[2n+2\gamma_{\kappa}+1-(N_{n\kappa}+N_{n'\kappa})(2\kappa+N_{n\kappa}-N_{n'\kappa})\right]
\nonumber \\
\times
\left\{
4(n+\gamma_{\kappa})(N_{n\kappa}+N_{n'\kappa})+2\epsilon_{n\kappa}\Delta_{n\kappa}^{(+)}+(N_{n'\kappa}-\kappa)\left[\epsilon_{n\kappa}(N_{n\kappa}+\kappa)+4\right]
\right\},
\label{4.31}
\end{eqnarray}
\begin{equation}
A_{n'}^{(5)}=-(N_{n\kappa}-\kappa)^2(n+2\gamma_{\kappa}+1)^2(n+2\gamma_{\kappa}+2)^2 \left[4(n+\gamma_{\kappa}-1)\epsilon_{n\kappa}+N_{n\kappa}+N_{n'\kappa}\right],
\label{4.32}
\end{equation}
where we have defined $\Delta_{n\kappa}^{(\pm)}=3n^2+6n\gamma_{\kappa}\pm3n\pm3\gamma_{\kappa}+2\gamma_{\kappa}^2+1$. After tedious calculations, the expression in Eq.\ (\ref{4.27}) may be cast into a much simpler form
\begin{eqnarray}
\mathcal{I}_{\kappa}^{(\infty)}&=&-\frac{\alpha a_0^4}{4Z^2} 
\left\{
(n+\gamma_{\kappa})
\left[
10(n+\gamma_{\kappa})^4+(13\kappa^2+10\gamma_{\kappa}^2)(n+\gamma_{\kappa})^2+N_{n\kappa}^2(6\kappa^2-3\gamma_{\kappa}^2+2)
\right]
N_{n\kappa} \right.
\nonumber \\
&& \quad-\left.2\kappa 
\left[(\kappa^2-\gamma_{\kappa}^2)\left(\kappa^2(n+\gamma_{\kappa})^2-(\gamma_{\kappa}^2-1)N_{n\kappa}^2\right)+(n+\gamma_{\kappa})^2\left(10N_{n\kappa}^2-8\kappa^2+2\gamma_{\kappa}^2+5\right)N_{n\kappa}^2
\right]
\right\}.
\nonumber \\
\label{4.33}
\end{eqnarray}
Combining Eqs.\ (\ref{4.24}), (\ref{4.26}) and (\ref{4.33}), as the formula in Eq.\ (\ref{4.18}) requires, we obtain
\begin{eqnarray}
\mathcal{Q}_{\kappa}^{(1)}=\frac{\alpha a_0^4}{Z^4} (4\pi \epsilon_0) c B  \frac{\kappa^2\mu(\kappa^2\!-\!\gamma_{\kappa}^2) (4\kappa^2\!-\!12\mu^2\!-\!1)}{2(4\kappa^2-1)^2} \!
\left\{(5n^2+10n \gamma_{\kappa}+4\gamma_{\kappa}^2+1)N_{n\kappa}^2-\kappa(n+\gamma_{\kappa})\!\left[2\kappa(n+\gamma_{\kappa})+3N_{n \kappa}\right]\right\}.
\nonumber \\
\label{4.34}
\end{eqnarray}

We turn now to the derivation of the expression for the two remaining components of $\mathcal{Q}^{(1)}$. In view of Eqs.\ (\ref{3.22}), (\ref{4.6}), (\ref{4.7}) and (\ref{4.9}), their sum may be written as
\begin{equation}
\mathcal{Q}_{-\kappa+1}^{(1)}+\mathcal{Q}_{-\kappa-1}^{(1)}=\frac{3(4\pi \epsilon_0) c B \mu}{2(4\kappa^2-1)} 
\sum_{\kappa'} \left[\frac{(2\kappa-1)^2-4\mu^2}{(2\kappa-1)(2\kappa-3)}\delta_{\kappa',-\kappa+1} 
+\frac{(2\kappa+1)^2-4\mu^2}{(2\kappa+1)(2\kappa+3)}\delta_{\kappa',-\kappa-1}\right] 
\mathcal{R}_{\kappa'},
\label{4.35}
\end{equation}
where we define
\begin{eqnarray}
\mathcal{R}_{\kappa'}=\sum_{n'=-\infty}^{\infty}\frac{1}{\mu_{n' \kappa'}^{(0)}-1}\int_0^{\infty} \textrm{d}r \: r^2 \left[P_{n\kappa}^{(0)}(r) S_{n' \kappa'}^{(0)}(r)+Q_{n\kappa}^{(0)}(r)T_{n'\kappa'}^{(0)}(r)\right]
\nonumber \\
\times \int_0^{\infty} \textrm{d}r' \: r' \left[\mu_{n' \kappa'}^{(0)} Q_{n\kappa}^{(0)}(r') S_{n' \kappa'}^{(0)}(r') + P_{n\kappa}^{(0)}(r') T_{n' \kappa'}^{(0)}(r') \right].
\label{4.36}
\end{eqnarray}
To evaluate the first radial integral on the right-hand side of above equation, we shall exploit Eqs. (\ref{4.10})--(\ref{4.11}) and (\ref{2.6})--(\ref{2.7}), with the Laguerre polynomials written in the form
\begin{equation}
L_n^{(\beta)}(\rho)=\sum_{k=0}^{n} \frac{(-)^k}{k!} \left( \begin{array}{c} n+\beta \\
n-k \end{array} \right) \rho^k,
\label{4.37}
\end{equation}
and transform the integration variable according to $x=\lambda_{n\kappa}r$. After these steps, we obtain
\begin{eqnarray}
\int_0^{\infty} \textrm{d}r \: r^2 \left[P_{n\kappa}^{(0)}(r) S_{n' \kappa'}^{(0)}(r)+Q_{n\kappa}^{(0)}(r) T_{n' \kappa'}^{(0)}(r) \right]=\frac{2 \tilde{f}_{n\kappa} \tilde{g}_{n\kappa}}{\lambda_{n\kappa}^3} \Gamma(n+2\gamma_{\kappa}) \sum_{k=0}^{n}\frac{(-)^k}{k!(n-k)!\Gamma(k+2\gamma_{\kappa}+1)}
\nonumber
\\
\times
\int_0^{\infty}\textrm{d}x \: x^{\gamma_{\kappa}+\gamma_{\kappa'}+k+2} \textrm{e}^{-x} \left[C_{k}^{(1)}L_{|n'|-1}^{(2\gamma_{\kappa'})}(x)+\frac{\kappa'-N_{n'\kappa'}}{|n'|+2\gamma_{\kappa'}}C_{k}^{(2)}L_{|n'|}^{(2\gamma_{\kappa'})}(x)
\right],
\label{4.38}
\end{eqnarray}
with
\begin{equation}
C_{k}^{(1)}=(n-k)+\epsilon_{n\kappa}\left(\kappa-N_{n\kappa}\right) \qquad \textrm{and} \qquad C_{k}^{(2)}=\epsilon_{n\kappa}(n-k)+\left(\kappa-N_{n\kappa}\right).
\label{4.39}
\end{equation}
Utilizing the following formula, Eq.\ (7.414.11) in Ref.\ \cite{Grad94},
 \begin{equation}
\int_0^{\infty} \textrm{d}\rho \: \rho^{\gamma}  \textrm{e}^{-\rho}  L_n^{(\beta)}(\rho)=\frac{\Gamma(\gamma+1) \Gamma(n+\beta-\gamma)}{n! \Gamma(\beta-\gamma)}=(-)^{n}\frac{\Gamma(\gamma+1)\Gamma(\gamma-\beta+1)}{n!\Gamma(\gamma-\beta-n+1)} \qquad \quad \left[\Real(\gamma)>-1 \right]
\label{4.40}
\end{equation}
and again the relation (\ref{4.14}), after some rearrangements we get
\begin{eqnarray}
\int_0^{\infty} \textrm{d}r \: r^2 \left[P_{n\kappa}^{(0)}(r) S_{n' \kappa'}^{(0)}(r)+Q_{n\kappa}^{(0)}(r) T_{n' \kappa'}^{(0)}(r) \right]=\frac{2 \tilde{f}_{n\kappa} \tilde{g}_{n\kappa}}{\lambda_{n\kappa}^3} \frac{\Gamma(n+2\gamma_{\kappa})}{(|n'|\!-\!1)!(N_{n'\kappa'}+\kappa')} \sum_{k=0}^{n}\frac{(-)^k\Gamma(\gamma_{\kappa}+\gamma_{\kappa'}+k+3)}{k!(n-k)!\Gamma(k+2\gamma_{\kappa}+1)} 
\nonumber \\
\times  \frac{\Gamma(|n'|+\gamma_{\kappa'}-\gamma_{\kappa}-k-3)}{\Gamma(\gamma_{\kappa'}-\gamma_{\kappa}-k-2)}
\left[
C_{k}^{(1)}(N_{n'\kappa'}+\kappa')-C_{k}^{(2)}(|n'|+\gamma_{\kappa'}-\gamma_{\kappa}-k-3)
\right].\quad
\label{4.41}
\end{eqnarray}
Proceeding in a similar way, one arrives at
\begin{eqnarray}
\int_0^{\infty} \textrm{d}r \: r \left[\mu_{n' \kappa'}^{(0)} Q_{n\kappa}^{(0)}(r) S_{n' \kappa'}^{(0)}(r)+P_{n\kappa}^{(0)}(r) T_{n' \kappa'}^{(0)}(r) \right]=\frac{\alpha Z}{N_{n\kappa}}\frac{\tilde{f}_{n\kappa} \tilde{g}_{n\kappa}}{\lambda_{n\kappa}^2} \frac{\Gamma(n+2\gamma_{\kappa})}{(|n'|\!-\!1)!(N_{n'\kappa'}+\kappa')} \sum_{p=0}^{n}\frac{(-)^p \Gamma(\gamma_{\kappa}+\gamma_{\kappa'}+p+2)}{p!(n-p)!\Gamma(p+2\gamma_{\kappa}+1)}
\nonumber
\\
\times  \frac{\Gamma(|n'|+\gamma_{\kappa'}-\gamma_{\kappa}-p-2)}{\Gamma(\gamma_{\kappa'}-\gamma_{\kappa}-p-1)}
 \Big\{(\mu_{n' \kappa'}^{(0)}+1) 
\left[(n-p)(N_{n'\kappa'}+\kappa')+(\kappa-N_{n \kappa})(|n'|+\gamma_{\kappa'}-\gamma_{\kappa}-p-2)\right] \quad 
\nonumber
\\
-(\mu_{n' \kappa'}^{(0)}-1)\left[(\kappa-N_{n \kappa})(N_{n'\kappa'}+\kappa')+(n-p)(|n'|+\gamma_{\kappa'}-\gamma_{\kappa}-p-2)\right]\Big\}. \quad \quad
\label{4.42}
\end{eqnarray}
Inserting Eqs.\ (\ref{2.9}), (\ref{4.12}), (\ref{4.13}), (\ref{4.41}), and (\ref{4.42}) into Eq.\ (\ref{4.36}), with some labor we obtain
\begin{eqnarray}
\mathcal{R}_{\kappa'}&=&\frac{\alpha a_0^4}{Z^4}\frac{n!\Gamma(n+2\gamma_{\kappa}+1)N_{n\kappa}^3}{64(N_{n\kappa}-\kappa)} \sum_{k=0}^{n} \sum_{p=0}^{n}  \frac{\mathcal{X}(k)\mathcal{Y}(p)}{\Gamma(\gamma_{\kappa'}-\gamma_{\kappa}-k-2)\Gamma(\gamma_{\kappa'}-\gamma_{\kappa}-p-1)} {}
\nonumber
\\
&&\times \sum_{n'=-\infty}^{\infty}\frac{\Gamma(|n'|+\gamma_{\kappa'}-\gamma_{\kappa}-k-3)\Gamma(|n'|+\gamma_{\kappa'}-\gamma_{\kappa}-p-2)}{|n'|!\Gamma(|n'|+2\gamma_{\kappa'}+1) (|n'|+\gamma_{\kappa'}-\gamma_{\kappa}-n)} \frac{\kappa'-N_{n' \kappa'}}{N_{n' \kappa'}}{}
\nonumber
\\
&&\times \left[C_{k}^{(1)}(N_{n'\kappa'}+\kappa')-C_{k}^{(2)}(|n'|+\gamma_{\kappa'}-\gamma_{\kappa}-k-3)\right]{}
\nonumber
\\
&&\times \left\{(|n'|+\gamma_{\kappa'}-\gamma_{\kappa}-n)\left[(\kappa-N_{n \kappa})(N_{n'\kappa'}+\kappa')+(n-p)(|n'|+\gamma_{\kappa'}-\gamma_{\kappa}-p-2)\right] \right.
\nonumber
\\
&&\quad-\left.(N_{n'\kappa'}+N_{n \kappa}) 
\left[(n-p)(N_{n'\kappa'}+\kappa')+(\kappa-N_{n \kappa})(|n'|+\gamma_{\kappa'}-\gamma_{\kappa}-p-2)\right]\right\},
\label{4.43}
\end{eqnarray}
where
\begin{equation}
\mathcal{X}(k)=\frac{(-)^{k}}{k!(n-k)!} \frac{\Gamma(\gamma_{\kappa}+\gamma_{\kappa'}+k+3)}{\Gamma(k+2\gamma_{\kappa}+1)} \quad \textrm{and} \quad \mathcal{Y}(p)=\frac{(-)^{p}}{p!(n-p)!} \frac{\Gamma(\gamma_{\kappa}+\gamma_{\kappa'}+p+2)}{\Gamma(p+2\gamma_{\kappa}+1)}.
\label{4.44}
\end{equation}
This result may be cast into another, much more perspicuous form, if in the series $\sum_{n'=-\infty}^{\infty}(\ldots)$ one collects together terms with the same absolute value of the summation index $n'$ (the Sturmian radial quantum number).  Proceeding in that way, after much labor, using Eq.\ (\ref{4.14}) and the extremely useful identity $\gamma_{\kappa'}^2-\gamma_{\kappa}^2=\kappa'^2-\kappa^2$, one finds that
\begin{eqnarray}
\mathcal{R}_{\kappa'}&=&\frac{\alpha a_0^4}{Z^4}\frac{n!\Gamma(n+2\gamma_{\kappa}+1)N_{n\kappa}^3}{32(N_{n\kappa}-\kappa)} \sum_{k=0}^{n} \sum_{p=0}^{n} \frac{\mathcal{X}(k)\mathcal{Y}(p)}{\Gamma(\gamma_{\kappa'}-\gamma_{\kappa}-k-2)\Gamma(\gamma_{\kappa'}-\gamma_{\kappa}-p-1)} {}
\nonumber
\\
&&\times \sum_{n'=0}^{\infty}\frac{\Gamma(n'+\gamma_{\kappa'}-\gamma_{\kappa}-k-3)\Gamma(n'+\gamma_{\kappa'}-\gamma_{\kappa}-p-2)}{n'!\Gamma(n'+2\gamma_{\kappa'}+1) (n'+\gamma_{\kappa'}-\gamma_{\kappa}-n)} {}
\nonumber
\\
&&\times 
\left\{
C_{k}^{(2)}(N_{n\kappa}-\kappa)(N_{n\kappa}-\kappa')(n'+\gamma_{\kappa'}-\gamma_{\kappa}-k-3)(n'+\gamma_{\kappa'}-\gamma_{\kappa}-p-2)\right.
\nonumber \\
&& \quad
+\left[C_{k}^{(1)}(n-p)(\kappa+\kappa')+2C_{k}^{(1)}(N_{n\kappa}-\kappa)-C_{k}^{(2)}(n-p)(n-k-3) \right]n'(n'+2\gamma_{\kappa'})
\nonumber \\
&& \quad
+\,C_{k}^{(2)}(n-p)(n'+\gamma_{\kappa'}-\gamma_{\kappa}-k-3)(n'+\gamma_{\kappa'}-\gamma_{\kappa}-p-2)(n'+\gamma_{\kappa'}-\gamma_{\kappa}-n)
\nonumber \\
&& \left. \quad 
-\,C_{k}^{(2)}(n-p)n'(n'+2\gamma_{\kappa'})(n'+\gamma_{\kappa'}-\gamma_{\kappa}-n)	
	\right\}.
\label{4.45}
\end{eqnarray}
It is possible to simplify the above formula. To achieve the first modification, we notice that the sum of the two series $\sum_{n'=0}^{\infty}(...)$ formed by using the last two components from the curly braces equals zero. Next, we may express the remaining two series in terms of the hypergeometric functions ${}_3F_2$ of the unit argument. Since it holds that \cite{Bail35,Slat66}  
\begin{eqnarray}
 {}_3F_2 \left(\begin{array}{c} 
a_1, a_2, a_3\\
b_1, b_2
\end{array};1 \right)=\frac{\Gamma(b_1)\Gamma(b_2)}{\Gamma(a_1) \Gamma(a_2)\Gamma(a_3)} \sum_{n=0}^{\infty} \frac{\Gamma(a_1+n)\Gamma(a_2+n)\Gamma(a_3+n)}{n!\Gamma(b_1+n)\Gamma(b_2+n)} \quad \qquad \left[\Real\left(b_1+b_2-a_1-a_2-a_3 \right)>0\right], 
\nonumber \\
\label{4.46} 
 \end{eqnarray}
Eq. (\ref{4.45}) becomes
\begin{eqnarray}
\mathcal{R}_{\kappa'}&=&\frac{\alpha a_0^4}{Z^4}\frac{n!\Gamma(n+2\gamma_{\kappa}+1)N_{n\kappa}^3}{32(N_{n\kappa}-\kappa)\Gamma(2\gamma_{\kappa'}+1)} \sum_{k=0}^{n} \sum_{p=0}^{n} \mathcal{X}(k)\mathcal{Y}(p) \left[
\frac{C_{k}^{(2)}(N_{n\kappa}-\kappa)(N_{n\kappa}-\kappa')}{\gamma_{\kappa'}-\gamma_{\kappa}-n} \right.
\nonumber
\\
&&\times
 {}_3F_2 \left(\begin{array}{c} 
\gamma_{\kappa'}-\gamma_{\kappa}-k-2, \gamma_{\kappa'}-\gamma_{\kappa}-p-1, \gamma_{\kappa'}-\gamma_{\kappa}-n\\
\gamma_{\kappa'}-\gamma_{\kappa}-n+1,2\gamma_{\kappa'}+1
\end{array};1 \right) 
\nonumber
\\
&&\quad +\frac{C_{k}^{(1)}(n-p)(\kappa+\kappa')+2C_{k}^{(1)}(N_{n\kappa}-\kappa)-C_{k}^{(2)}(n-p)(n-k-3)}{\gamma_{\kappa'}-\gamma_{\kappa}-n+1}{}
\nonumber
\\
&&\quad \left.\times {}_3F_2 \left(\begin{array}{c} 
\gamma_{\kappa'}-\gamma_{\kappa}-k-2, \gamma_{\kappa'}-\gamma_{\kappa}-p-1, \gamma_{\kappa'}-\gamma_{\kappa}-n+1\\
\gamma_{\kappa'}-\gamma_{\kappa}-n+2,2\gamma_{\kappa'}+1
\end{array};1 \right)
\right].
\label{4.47}
\end{eqnarray}
The first ${}_3F_{2}$ function may be eliminated with the help of the recurrence formula
\begin{eqnarray}{}_3F_2 \left( \begin{array}{c} 
a_1, a_2, a_3-1\\
a_3,b
\end{array};1 \right)=-\frac{(a_1-a_3)(a_2-a_3)}{a_3(b-a_3)} {}_3F_2 \left( \begin{array}{c} 
a_1, a_2, a_3\\
a_3+1,b
\end{array};1 \right)
+\frac{\Gamma(b)\Gamma(b-a_1-a_2+1)}{(b-a_3)\Gamma(b-a_1)\Gamma(b-a_2)}
\nonumber \\
\left[\Real(b-a_1-a_2)>-1 \right]
\label{4.48}
\end{eqnarray}
and therefore Eq.\ (\ref{4.47}) can be rewritten as
\begin{eqnarray}
\mathcal{R}_{\kappa'}&=&\frac{\alpha a_0^4}{Z^4} \frac{N_{n \kappa}^3}{32(N_{n \kappa}+\kappa')}\Bigg[\mathcal{W}_{n\kappa}+ \frac{n! \Gamma(n+2\gamma_{\kappa}+1)}{(N_{n\kappa}-\kappa)(\gamma_{\kappa'}-\gamma_{\kappa}-n+1)\Gamma(2\gamma_{\kappa'}+1)} 
\nonumber
\\
&&\times \sum_{k=0}^n\sum_{p=0}^n \tilde{\mathcal{X}}(k) \tilde{\mathcal{Y}}(p) {}_3F_2 \left(\begin{array}{c} 
\gamma_{\kappa'}-\gamma_{\kappa}-k-2, \gamma_{\kappa'}-\gamma_{\kappa}-p-1, \gamma_{\kappa'}-\gamma_{\kappa}-n+1\\
\gamma_{\kappa'}-\gamma_{\kappa}-n+2,2\gamma_{\kappa'}+1
\end{array};1 \right) \Bigg],
\label{4.49}
\end{eqnarray}
where
\begin{equation}
\tilde{\mathcal{X}}(k)=\left[C_{k}^{(1)}(N_{n\kappa}+\kappa')-C_{k}^{(2)}(n-k-3) \right]\mathcal{X}(k),
\label{4.50}
\end{equation}
\begin{equation}
\tilde{\mathcal{Y}}(p)=\left[(n-p)(\kappa+\kappa')+2(N_{n\kappa}-\kappa) \right]\mathcal{Y}(p)
\label{4.51}
\end{equation}
and
\begin{equation}
\mathcal{W}_{n\kappa}=n!\Gamma(n+2\gamma_{\kappa}+1) 
\sum_{k=0}^n\sum_{p=0}^n\frac{(-)^{k+p+1}\Gamma(2\gamma_{\kappa}+k+p+5) C_{k}^{(2)}}{k!p!(n-p)!(n-k)!\Gamma(k+2\gamma_{\kappa}+1)\Gamma(p+2\gamma_{\kappa}+1)}.
\label{4.52}
\end{equation}
The above expression for $\mathcal{W}_{n\kappa}$ has a very similar form to the formula defined in Eq. (A.1) in Ref.\ \cite{Stef15a}. Basing on the analysis carried out in the Appendix to that article, after some further straightforward calculations, Eq.\ (\ref{4.52}) may be cast to the form
\begin{eqnarray}
\mathcal{W}_{n\kappa}=\frac{2(N_{n\kappa}-\kappa)}{N_{n\kappa}} 
\left\{
4\kappa (n+\gamma_{\kappa})^2\left[7(n+\gamma_{\kappa})^2-3\gamma_{\kappa}^2+5\right]
+N_{n\kappa} \left[63(n+\gamma_{\kappa})^4+70(n+\gamma_{\kappa})^3\right. \right.
\nonumber \\
\left. \left.-21(2\gamma_{\kappa}^2-5)(n+\gamma_{\kappa})^2 -10(3\gamma_{\kappa}^2-5)(n+\gamma_{\kappa})+3(\gamma_{\kappa}^2-1)(\gamma_{\kappa}^2-4) \right]
\right\}.
\label{4.53}
\end{eqnarray}
If Eq.\ (\ref{4.49}) is inserted into Eq.\ (\ref{4.35}), the sum of the two considered components of $\mathcal{Q}^{(1)}$ is
\begin{eqnarray}
\mathcal{Q}_{-\kappa+1}^{(1)}+\mathcal{Q}_{-\kappa-1}^{(1)}=\frac{3}{64}\frac{\alpha a_0^4}{Z^4} (4\pi \epsilon_0) c B \mu \frac{N_{n\kappa}^3}{(4\kappa^2-1)^2} \sum_{\kappa'} \frac{\eta_{\kappa \mu}^{(+)}\delta_{\kappa',-\kappa+1}+\eta_{\kappa \mu}^{(-)}\delta_{\kappa',-\kappa-1}}{N_{n\kappa}+\kappa'}
\nonumber \\
\times
\left[
\mathcal{W}_{n\kappa}
+\frac{n!\Gamma(n+2\gamma_{\kappa}+1)}{(N_{n\kappa}-\kappa)(\gamma_{\kappa'}-\gamma_{\kappa}-n+1)\Gamma(2\gamma_{\kappa'}+1)} \sum_{k=0}^n\sum_{p=0}^n \tilde{\mathcal{X}}(k) \tilde{\mathcal{Y}}(p) \right.
\nonumber \\
\times \left. 
 {}_3F_2 \left(\begin{array}{c} 
\gamma_{\kappa'}-\gamma_{\kappa}-k-2,\:
\gamma_{\kappa'}-\gamma_{\kappa}-p-1,\: 
\gamma_{\kappa'}-\gamma_{\kappa}-n+1 \\
\gamma_{\kappa'}-\gamma_{\kappa}-n+2,\:
2\gamma_{\kappa'}+1
\end{array}
;1 
\right)
\right],
\label{4.54}
\end{eqnarray} 
where $\eta_{\kappa \mu}^{(\pm)}=[(2\kappa \mp 1)^2-4\mu^2](2\kappa \pm 1)/(2\kappa \mp 3)$. Finally, putting Eqs.\ (\ref{4.34}) and (\ref{4.54}) into Eq.\ (\ref{4.9}), and utilizing also Eq. (\ref{4.53}), we find that the electric quadrupole moment induced by a weak dipole magnetic field in the Dirac one-electron atom in the state characterized by the set of quantum numbers $\{n, \kappa, \mu\}$, is given by
\begin{eqnarray}
\mathcal{Q}^{(1)} \equiv \mathcal{Q}_{n \kappa \mu}^{(1)}&=&\frac{\alpha a_0^4}{Z^4}  \frac{(4\pi \epsilon_0) c B \mu}{64(4\kappa^2-1)^2} 
\left\{
\Theta_{n\kappa \mu}^{(\texttt{I})}+\sum_{\kappa'} \frac{\eta_{\kappa \mu}^{(+)}\delta_{\kappa',-\kappa+1}+\eta_{\kappa \mu}^{(-)}\delta_{\kappa',-\kappa-1}}{N_{n\kappa}+\kappa'} \right.
\nonumber \\
&& \times\left[
\Theta_{n\kappa}^{(\texttt{II})}+\frac{3n!\Gamma(n+2\gamma_{\kappa}+1)N_{n\kappa}^3}{(N_{n\kappa}-\kappa)(\gamma_{\kappa'}-\gamma_{\kappa}-n+1)\Gamma(2\gamma_{\kappa'}+1)} \sum_{k=0}^n\sum_{p=0}^n \tilde{\mathcal{X}}(k) \tilde{\mathcal{Y}}(p) \right.
\nonumber \\
&& \quad \times \left. \left.
 {}_3F_2 \left(\begin{array}{c} 
\gamma_{\kappa'}-\gamma_{\kappa}-k-2,\:
\gamma_{\kappa'}-\gamma_{\kappa}-p-1,\: 
\gamma_{\kappa'}-\gamma_{\kappa}-n+1 \\
\gamma_{\kappa'}-\gamma_{\kappa}-n+2,\:
2\gamma_{\kappa'}+1
\end{array}
;1 
\right)
\right]
\right\},
\label{4.55}
\end{eqnarray}
with
\begin{eqnarray}
\Theta_{n\kappa \mu}^{(\texttt{I})}=32\kappa^2(\kappa^2-\mu^2)(4\kappa^2-12\mu^2-1) 
\left\{(5n^2+10n \gamma_{\kappa}+4\gamma_{\kappa}^2+1)N_{n\kappa}^2-\kappa(n+\gamma_{\kappa})\left[2\kappa(n+\gamma_{\kappa})+3N_{n \kappa}\right] \right.\}
\label{4.56}
\end{eqnarray}
and
\begin{eqnarray}
\Theta_{n\kappa}^{(\texttt{II})}=6(N_{n\kappa}-\kappa)N_{n\kappa}^2 \left\{
4\kappa (n+\gamma_{\kappa})^2\left[7(n+\gamma_{\kappa})^2-3\gamma_{\kappa}^2+5\right]+\left[63(n+\gamma_{\kappa})^4+70(n+\gamma_{\kappa})^3\right. \right.
\nonumber \\
\left. \left. -21(2\gamma_{\kappa}^2-5)(n+\gamma_{\kappa})^2-10(3\gamma_{\kappa}^2-5)(n+\gamma_{\kappa})+3(\gamma_{\kappa}^2-1)(\gamma_{\kappa}^2-4) \right]N_{n\kappa} 
\right\}.
\label{4.57}
\end{eqnarray}
With the above result, we are able to obtain values of the induce electric quadrupole moment in any discrete energy state of the atom.   

As every new results, the expression for $\mathcal{Q}_{n \kappa \mu}^{(1)}$ should be subjected to some kind of verification. The only test, which allows one to support the correctness of our formula is to check its form for some particular states of the atom and compare the resulting formula with other results available in the literature. To accomplish the goal, we shall insert $n=0$, $\kappa=-1$ and $\mu=\pm 1/2$ into Eq.\ (\ref{4.55}), arriving at the expression for the electric quadrupole moment induced in the \emph{ground} state of the relativistic hydrogenlike atom, which is 
\begin{eqnarray}
\mathcal{Q}_{0,-1,\pm \frac{1}{2}}^{(1)} \equiv \mathcal{Q}_{g}^{(1)}= \textrm{sgn}(\mu) \frac{\alpha a_0^4}{Z^4} \frac{(4\pi \epsilon_0) c B}{360} \frac{\Gamma(2\gamma_1+5)}{\Gamma(2\gamma_1+1)} \left[1-\frac{6(\gamma_1+1)\Gamma(\gamma_1+\gamma_2+2)\Gamma(\gamma_1+\gamma_2+3)}{(\gamma_2-\gamma_1+1) \Gamma(2\gamma_1+5)\Gamma(2\gamma_2+1)} \right.
\nonumber
\\
 \times 
\left.
{}_3F_2 
\left( 
\begin{array}{c} 
\gamma_2-\gamma_1-2,\: 
\gamma_2-\gamma_1-1,\: 
\gamma_2-\gamma_1+1 \\
\gamma_2-\gamma_1+2,\:  
2\gamma_2+1
\end{array}
;1 
\right) 
\right].
\label{4.58}
\end{eqnarray}
Transforming the hypergeometric function with the aid of the following formula:
\begin{eqnarray}
{}_3F_2 \left( \begin{array}{c} 
a_1, a_2, a_3\\
a_3+1,b
\end{array};1 \right)=-\frac{a_3(b-a_3)}{(a_1-a_3)(a_2-a_3)}{}_3F_2 \left( \begin{array}{c} 
a_1, a_2, a_3-1\\
a_3,b
\end{array};1 \right) +\frac{a_3 \Gamma(b)\Gamma(b-a_1-a_2+1)}{(a_1-a_3)(a_2-a_3)\Gamma(b-a_1)\Gamma(b-a_2)} {}
\nonumber
\\
\left[\Real(b-a_1-a_2)>-1 \right], \quad
\label{4.59}
\end{eqnarray}
Eq.\ (\ref{4.58}) can be rewritten as
\begin{eqnarray}
\mathcal{Q}_{g}^{(1)}&=&\textrm{sgn}(\mu) \frac{\alpha^2 e a_0^2}{Z^4}\frac{B}{b_0} \frac{\Gamma(2\gamma_1+5)}{1440\Gamma(2\gamma_1)} \left[-1+\frac{(\gamma_1+1)(\gamma_1+\gamma_2)\Gamma(\gamma_1+\gamma_2+2)\Gamma(\gamma_1+\gamma_2+3)}{\gamma_1\Gamma(2\gamma_1+5) \Gamma(2\gamma_2+1)} \right.
\nonumber \\
&&  \times 
\left.
{}_3F_2 
\left( 
\begin{array}{c} 
\gamma_2-\gamma_1-2,\: 
\gamma_2-\gamma_1-1,\:
\gamma_2-\gamma_1 \\
\gamma_2-\gamma_1+1,\: 
2\gamma_2+1
\end{array}
;1 
\right) 
\right],
\label{4.60}
\end{eqnarray}
where 
\begin{equation}
b_0=\frac{\mu_0}{4\pi} \frac{\mu_B}{a_0^3}=\frac{\alpha^2 \hbar}{2 e a_0^2} \simeq 6.26 \: \textrm{T}
\label{4.61}
\end{equation}
is the atomic unit of the magnetic induction ($\mu_0$ is the vacuum permeability and $\mu_B$ is the Bohr magneton). The above result, \emph{linear} in the perturbing field strength $B$, is identical to the corresponding formula obtained by us some time ago by direct calculations \cite{Szmy12}.

It is worthwhile to recall here one of the most interesting concluding remarks from the above-mentioned article, i.e., that in the nonrelativistic limit (when $\gamma_{\kappa} \to |\kappa|$) the expression from Eq. (\ref{4.60}) tends to zero. This confirms the earlier results  of calculations of that quantity for the one-electron atom, obtained on the basis of the \emph{nonrelativistic} theories \cite{Coul56,Turb87,Pote01}, in which the leading term in the expansion of the induced electric quadrupole moment in powers of the field strength is \emph{quadratic} in $B$.

\section{Conclusions}
\label{V}
\setcounter{equation}{0}

In this work, we have analyzed the electric multipole moments induced in the relativistic hydrogenlike atom in an arbitrary discrete energy eigenstate by a weak, uniform, static magnetic field. We have shown that, to the first order in the perturbing field, only even-order electric multipole moments can be induced in the system. Next, we have derived analytically a closed-form expression for the induced electric quadrupole moment for any state of the Dirac one-electron atom. The result has the form of a double finite sum involving the generalized hypergeometric functions ${}_3F_{2}$ of the unit argument; for the atomic ground state it reduces to the formula for the considered quantity found by us some time ago \cite{Szmy12}. 

We have discussed this physical effect on the basis of the relativistic theory and in a general overview. There are some articles \cite{Coul56, Turb87, Pote01}, in which this physical problem is treated in a nonrelativistic manner, but they concern only the atomic ground state. The result presented in this work not only has been obtained by taking into account the relativity, but also generalizes the existing formula for $\mathcal{Q}^{(1)}(B)$ to an arbitrary state of the atom.

The calculations of the actual value of the induced electric quadrupole moment that we have carried out in Sec.\ \ref{IV} provide another example of the usefulness of the Sturmian expansion of the generalized Dirac--Coulomb Green function \cite{Szmy97} for analytical determination of electromagnetic properties of the relativistic hydrogenlike atom in an arbitrary discrete energy eigenstate.

\begin{acknowledgments}
I am grateful to Professor R.\ Szmytkowski for his support during the preparation of this work and for commenting on the manuscript.
\end{acknowledgments}

\end{document}